\documentclass[a4paper,fleqn]{cas-sc}

\usepackage[authoryear]{natbib}
\usepackage{ulem}
\usepackage{threeparttable}
\usepackage{pifont}
\usepackage[left]{lineno}

\begin{document}
\let\WriteBookmarks\relax
\def\floatpagepagefraction{1}
\def\textpagefraction{.001}
\shorttitle{Dust, gas activity and morphology in comet 12P/Pons-Brooks}
\shortauthors{M. Hus\'arik et~al.}

\title [mode = title]{Dust, gas activity and morphology of comet 12P/Pons-Brooks at heliocentric distances beyond 1.1 au }                      



\author[1]{Marek Hus\'arik}[type=editor,
                        auid=000,bioid=1,
                        orcid=0000-0002-5932-7214]
\cormark[1]
\fnmark[1]
\ead{mhusarik@ta3.sk}
\credit{Conceptualization, Data curation, Methodology, Writing - Original draft preparation}
\affiliation[1]{organization={Astronomical Institute of the Slovak Academy of Sciences},
                city={Tatransk\'{a} Lomnica},
                postcode={05960}, 
                country={Slovak Republic}}

\author[2]{Gulchehra Kokhirova}[orcid=0000-0001-8957-2061]
\credit{Data curation, Writing - Original draft preparation}
\affiliation[2]{organization={Institute of Astrophysics, National Academy of Sciences of Tajikistan},
	addressline={Ayni 299/5}, 
	postcode={734063}, 
	postcodesep={}, 
	city={Dushanbe},
	country={Republic of Tajikistan}}

\author[3]{Valerii Kleshchonok}[orcid=0000-0002-4832-371X
   ]

\credit{Investigation, Formal analysis, Writing - Original draft preparation}

\affiliation[3]{organization={Institut f\"{u}r Geophysik und Extraterrestrische Physik, Technische Universit\"{a}t Braunschweig},
                addressline={Mendelssohnstr. 3}, 
                postcodesep={D-38106}, 
                city={Braunschweig},
                country={Germany}}

\author[4]{K. Aravind}[orcid=0000-0002-8328-5667]
\credit{Investigation, Formal analysis}
\affiliation[4]{organization={Space Sciences, Technologies \& Astrophysics Research (STAR) Institute, University of Li\`{e}ge},
                postcodesep={B-4000}, 
                city={Li\`{e}ge},
                country={Belgium}}

\author[2]{Firuza Rakhmatullaeva}
\credit{Data curation, Investigation, Validation}

\author[5,6]{Margarita Safonova}[orcid=0000-0003-4893-6150]
\credit{Data curation}
\affiliation[5]{organization={M. P. Birla Institute of Fundamental Research, Bangalore},
                city={Bangalore},
                country={India}}
\affiliation[6]{organization={UNEX EuroMoonMars EuroSpaceHub},
                city={Leiden},
                country={The Netherlands}}

\author[1]{Oleksandra Ivanova}[orcid=0000-0001-7285-373X]
\credit{Formal analysis, Validation, Writing – original draft}

\author[1]{Olena Shubina}[orcid=0000-0002-3782-1262]
\credit{Validation, Writing – review \& editing}

\author[7,8]{Arsenii Kasianchuk}
\credit{Investigation}
\affiliation[7]{organization={Astronomical Observatory of Taras Shevchenko National University of Kyiv},
                addressline={Observatorna str. 3}, 
                postcode={04053}, 
                city={Kyiv},
                country={Ukraine}}
\affiliation[8]{organization={Main Astronomical Observatory on the National Academy of Sciences of Ukraine},
                addressline={27 Akademika Zabolotnoho St.}, 
                postcode={03143}, 
                city={Kyiv},
                country={Ukraine}}
    
\cortext[cor1]{Corresponding author}


\begin{abstract}
Comet 12P/Pons-Brooks is a periodic comet with an orbital period of approximately 71 years. Because of the period duration, aphelion of 35.3\,au, and highly inclined orbit $(74.2^\circ)$, it is classified as a Halley-type comet.

In this work, we present the results of the analysis of photometric and spectral observations of comet 12P/Pons-Brooks while it was at a heliocentric distance beyond 1.1\,au, before perihelion passage. Quasisynchronous observations were carried out from July 2023 to March 2024 using 0.61-m and 1.3-m telescopes at the Skalnat\'{e} Pleso Observatory, 1-m telescope Zeiss-1000 Sanglokh International Astronomical Observatory, and 0.7-m telescope AZT8 at the observation station of Taras Shevchenko National University of Kyiv. Photometric observations were conducted using B, V, and R filters of the Johnson-Cousins and Bessel photometric systems. We completed our data collection with broadband photometric and spectral observations on the 2-m HCT telescope at the Indian Astronomical Observatory.

During this period, the dust coma morphology and photometric data measured in the R filter, such as the apparent, absolute magnitudes and dust activity level, indicate several repeated outbursts in brightness and dust activity occurred around July 22, September 25, October 17 -- 23, and a significant outburst was recorded on November 1 -- 7. We detected strong gas emission features in the cometary spectrum belonging to CN, C$_{2}$, and C$_{3}$ molecules. Furthermore, gas production rates of these molecules were estimated using Haser's model. A comparison of dust activity level and CN gas production rate shows that dust activity increased significantly as the comet approached the Sun, while gas activity exhibited more moderate changes.
\end{abstract}



\maketitle

\section{Introduction}
Comet 12P/Pons-Brooks (hereafter 12P) is a periodic comet with an orbital period of approximately 71 years; it belongs to the Halley-type comet family. Comet 12P is one of the brightest known periodic comets, reaching an absolute visual magnitude of about 5 at the perihelion\footnote{\url{https://www.minorplanetcenter.net}}. It was discovered in 1812 by Jean-Louis Pons at the Marseille Observatory and was rediscovered by William Robert Brooks in 1883 \citep{1986AJ.....91..971Y}. Comet 12P exhibits orbital parameters characteristic of short-period comets, with an aphelion extending beyond Neptune's orbit at a distance of 33.6\,au. Orbital parameters of the comet taken from the NASA JPL database\footnote{\url{https://ssd.jpl.nasa.gov/tools/sbdb_lookup.html\#/?sstr=12P}} are listed in Table\,\ref{table_osculatelements}. The comet's nucleus is estimated to have a upper-limit diameter of $34 \pm 12$\,km \citep{2020RNAAS...4..101Y} and period of rotation $57 \pm 1$\,hr \citep{2024ATel16508....1K}, indicating its significant size relative to other studied comets. The comet is considered the potential progenitor of the weak December $\kappa$ Draconids meteor shower, which is active from late November to mid-December \citep{2016A&A...592A.107T}. The comet passed its perihelion on April 21, 2024. Comet 12P presents frequent brightness outbursts. One of the most significant events occurred in 1954 when the comet's brightness increased from magnitude 18 to 13 \citep{1972IAUS...45.....C}. More recently, observations during 2023 -- 2024 have shown repeated outbursts. The comet has started its activity at 11\,au \citep{2020RNAAS...4..101Y}. On July 20, 2023, a major outburst was observed, with the comet brightening from magnitude 17 to 12. This event was refined to July 19.57, followed by a smaller outburst on July 20.83 \citep{2023ATel16202....1M,2023ATel16194....1M}. Since then, several additional outbursts have been reported, including on September 4, 2023 \citep{2023ATel16229....1U}, September 22–25, 2023 \citep{2023ATel16254....1K}, October 5, 2023 \citep{2023ATel16270....1U}, November 14, 2023 \citep{2023ATel16282....1J}, December 12–14, 2023 \citep{2024ATel16408....1J}, and February 29, 2024 \citep{2024ATel16498....1J}. On July 25, 2023, comet 12P had colours measured as $B-R = 0.9 \pm 0.1$ and $V-R = 0.47 \pm 0.05$ \citep{2023ATel16315....1B}. The authors estimated the phase-corrected $Af\rho$ parameter in the R-band to correspond to $6535 \pm 349$\,cm, obtained with a photometric aperture radius of 4.1 arcseconds. 
Recent observations of comet 12P conducted with the TRAPPIST-North telescope revealed significant changes in dust and gas activity depending on the heliocentric distance \citep{2023ATel16282....1J}. The first outburst was recorded between September 22 and 25, when the comet was at a heliocentric distance of $3.20 - 3.16$\,au. During this period, the $A(0)f\rho$ value in the R filter nearly doubled (from 798\,cm to 1305\,cm), while the production rates of CN $(1.18 \times 10^{25} \mathrm{s}^{-1}$ to $1.20 \times 10^{25} \mathrm{s}^{-1})$ and C$_2$ $(0.84 \times 10^{25} \mathrm{s}^{-1}$ to $0.87 \times 10^{25} \mathrm{s}^{-1})$ showed minimal variation. The second major outburst occurred between October 3 and 7, at a heliocentric distance of $3.08 - 3.03$\,au. The $A(0)f\rho$ value in the RC filter increased sixfold (from 1285\,cm to 6715\,cm), while the CN production rate rose from $1.50 \times 10^{25} \mathrm{s}^{-1}$ to $1.91 \times 10^{25} \mathrm{s}^{-1}$, and the C$_2$ rate increased from $1.27 \times 10^{25} \mathrm{s}^{-1}$ to $2.47 \times 10^{25} \mathrm{s}^{-1}$. These observations highlight that dust activity increased significantly as the comet approached the Sun, while gas activity exhibited more moderate changes.
Spectral analysis of the observed data revealed the typical emission bands from neutral molecular species such as C$_2$, NH$_2$, and CN, along with a tentative detection of H$_2$O$^+$ \citep{2024EPSC...17..951V,2024EPSC...17.1247M}. Using the IRAM-30m and NOEMA radio telescopes in April 2024, molecules and isotopologues were identified in comet 12P/Pons-Brooks, including HCN, HNC, HC$_3$N, CH$_3$CN, HNCO, NH$_2$CHO, CH$_3$OH, H$_2$CO, CO, HCO$^+$, CS, H$_2$S \citep{2024EPSC...17..371B}. The water production rate of 12P peaked at $1.7 \times 10^{30}$\,s$^{-1}$ seven days before perihelion (April 21, 2024), showing a strong correlation with heliocentric distance \citep{2025PSJ.....6..306C}. The CN morphology exhibited two apparently face-on spiral features, roughly 180$^{\circ}$ apart, across all observed epochs \citep{2024DPS....5640107S,2024ATel16508....1K}.


   \begin{table*}
   \caption{Osculating orbital parameters of the comet 12P/Pons-Brooks (J2000.0).}             
   \label{table_osculatelements}      
   \centering          
   \begin{threeparttable}
   \begin{tabular}{cccccccc}    
\toprule
$a^{1}$,\,au & $e^{2}$ & $i^{3}$, deg & $q^{4}$,\,au & $P^{5}$,\,years & $T_{\mathrm Jup}$$^{6}$ & $t_{peri}$$^{7}$\\
\midrule
17.1848 & 0.9545 & 74.1908 & 0.7808 & 71.2405 & 0.598 & 2024-Apr-21.1\\
\bottomrule
   \end{tabular}
   \begin{tablenotes}
    \item Notes: ({\it 1}) semi-major axis, ({\it 2}) eccentricity, ({\it 3}) inclination, ({\it 4}) perihelion distance, ({\it 5}) orbital period, ({\it 6}) Jupiter Tisserand invariant, and ({\it 7}) time of perihelion passage.
   \end{tablenotes}
   \end{threeparttable}
   \end{table*}

In this work, we present the results of an analysis of photometric and spectral observations of comet 12P during its passage beyond 1.1\,au, before perihelion. The paper is organised as follows. Section 2 describes the observation technique and data processing, and Section 3 presents the results of the observations and their analysis, along with quantitative interpretation.  Finally, Section 4 provides discussion and conclusions.

\section{Observations and data processing}
\subsection{0.61-m telescope Skalnat\'e Pleso Observatory (61SP)}
Photometric observations of 12P were carried out from July to August 2023, utilising a 0.61-m Newtonian reflecting telescope located at the Skalnat\'e Pleso Observatory (Astronomical Institute of the Slovak Academy of Sciences, MPC code 056). The telescope was equipped with a CCD camera SBIG ST-10XME with a pixel size of $6.8\,\mu$m. The images were acquired with $2 \times 2$ binning, providing a field of view of $19^{\prime} \times 13^{\prime}$ and a pixel scale of $1.069^{\prime \prime}$/pixel. Observations were conducted using B, V, and R filters of the Johnson-Cousins photometric system, centred at wavelengths $\lambda_B = 0.44$\,$\mu$m (FWHM = 0.1\,$\mu$m), $\lambda_V = 0.53$\,$\mu$m (FWHM = 0.08$\,\mu$m), and $\lambda_R = 0.65$\,$\mu$m (FWHM = 0.13\,$\mu$m), respectively.

\subsection{1.3-m telescope Skalnat\'e Pleso Observatory (1.3SP)}
1.3SP was equipped with a camera FLI PL-230 with pixel sizes of 15\,$\mu$m $\times$ 15\,$\mu$m, a field of view was $9.8^{\prime} \times 9.8^{\prime}$  (pixel scale is $0.58^{\prime \prime}$/px), and $2 \times 2$ binning was used. Photometry with this camera is performed using B, V, and R filters of the Bessel system centred at $\lambda_B = 0.435$\,$\mu$m (FWHM = 0.078\,$\mu$m), $\lambda_V = 0.548$\,$\mu$m (FWHM = 0.099\,$\mu$m), and $\lambda_R = 0.635$\,$\mu$m (FWHM = 0.106\,$\mu$m), respectively.

\subsection{1-m telescope Zeiss-1000 (1.0SIAO)}
Photometric observations of the comet were conducted at the Sanglokh International Astronomical Observatory (Institute of Astrophysics, National Academy of Sciences of Tajikistan, IAU code 193) using the Zeiss-1000 telescope in 2023, over 5 nights: one in September, two in October, and two in November. The telescope's focal length is $f$ = 13.3\,m, resulting in an image scale of $63\,\mu$m/arcsec. The comet was imaged using a CCD camera FLI ProLine 16803. The camera's sensor size and field of view are $4096 \times 4096$ pixels, and $10^{\prime} \times 10^{\prime}$, respectively, with a scale of $0.579^{\prime \prime}$/px. To reduce data redundancy and enhance the signal-to-noise ratio, the images were recorded with $4 \times 4$ binning. The image quality, measured as the average FWHM of several stars from individual frames, ranged from $1.5^{\prime \prime}$ to $2.0^{\prime \prime}$. 
Multicolour BVR observations were performed using standard Johnson-Cousins broadband filters centred at $\lambda_B = 0.44$\,$\mu$m (FWHM = 0.1\,$\mu$m), $\lambda_V = 0.53$\,$\mu$m (FWHM = 0.08\,$\mu$m), $\lambda_R = 0.65$\,$\mu$m (FWHM = 0.13\,$\mu$m).

\subsection{0.7-m telescope AZT8 (AZT8 LSKU)}
Observations of the comet on March 10, 2024, were conducted at the observation station of Taras Shevchenko National University of Kyiv (IAU code 585), located in Lisnyky, using the telescope AZT8 $(D = 70\,\mathrm{cm}, f = 280\,\mathrm{cm}$. The C4-16000 camera with UBVR filters, mounted at the telescope's prime focus, served as the receiver. Due to weather conditions, we used images obtained in the R filter; twenty 20-second exposures were combined and used for processing.

\subsection{2-m telescope HCT Indian Astronomical Observatory (2.0HCT)}
Photometric (November 21, 2023) and spectroscopic (November 22, 2023) observations were obtained using the Himalayan Faint Object Spectrograph and Camera (HFOSC) mounted on the 2.0-m Himalayan Chandra Telescope (HCT) of the Indian Astronomical Observatory (IAO, IAU code N50). HFOSC is equipped with a Thompson CCD of $2048\times4096$ pixels with a pixel scale of 0.296$^{\prime \prime}$/px, equivalent to a total field of view (FOV) of $10^{\prime} \times 10^{\prime}$. The readout noise, gain and readout time of the CCD are 4.87\,e, 1.22\,e/ADU and 90 sec, respectively. Observations were done in the Keystone mode, which allows non-sidereal tracking of the comet. Photometry was performed in the broadband Sloan Digital Sky Survey (SDSS) $r^{\prime}$ filter with variable exposures. Spectroscopy was carried out using the Gr7/167l configuration, with a bandpass of 3700--6850~\AA, a resolution of $\lambda/\Delta\lambda = 1330$ and a slit width of 1.92$^{\prime\prime}$. Spectroscopy was performed in a single 900-sec exposure with the slit centred on the comet's nucleus and West--East oriented. To avoid identifying any telluric line as of cometary origin, a separate 900-sec sky frame was obtained with the telescope moved about 1 degree off the photocenter. Standard star, Feige110 (DOp type), was observed for flux calibration. Halogen lamp spectra, zero exposure frames, and FeAr lamp spectra were obtained for flat-fielding, bias subtraction, and wavelength calibration, respectively. 

{\it Data reduction} for observations obtained at 61SP, 1.3SP, 1.0SIAO, and AZT8LSKU followed standard procedures, including bias and dark subtraction, flat-field correction, and the removal of cosmic-ray artefacts, performed with custom IDL (Interactive Data Language) routines as described in previous studies (e.\,g., \citet{2016P&SS..121...10I,2017Icar..284..167I}). The background sky level was determined from coma- and star-free regions in the images, and only data obtained under photometric conditions were used. Absolute flux calibration was achieved using field stars from the UCAC4 catalogue (the United States Naval Observatory CCD Astrograph Catalog), which reports magnitude uncertainties of approximately $0.05-0.1$\,mag \citep{2013AJ....145...44Z}. The photometric data obtained with 2.0HCT were subjected to the usual image reduction process (bias subtraction, flat-fielding, and cosmic rays removal) using Image Reduction and Analysis Facility (IRAF) scripts developed by our group. Flat fields were constructed by taking the median of multiple dithered twilight-sky images, thereby removing contamination from cosmic rays and stellar sources. The comet spectrum was reduced and calibrated using custom Python scripts in combination with IRAF, following standard long-slit reduction procedures described in \cite{2021MNRAS.502.3491A}. The comet's spectrum was extracted from the reduced frame using Python routines. The standard-star spectra were used to determine the instrument's characteristic spectral trace, which, with necessary adjustments, was applied to trace and extract the comet spectra along the dispersion axis. The standard-star spectra were extracted with the IRAF task \textit{apall}, which provides effective sky subtraction using regions on both sides of the target. Wavelength and flux calibrations were performed using standard IRAF packages. A suitably scaled and slope-corrected solar spectrum \citep[see][for details]{2022Icar..38315042A} was applied to remove the reflected solar continuum contribution.

A summary of the observational details is presented in Tab.\,\ref{table_logobservations}. Notably, all data were collected before perihelion.

\begin{table*}
   \caption{Log of observations of comet 12P: the observation date and range of exposure time (in Coordinated Universal Time), heliocentric $(r)$ and geocentric $(\Delta)$ distances, phase angle $(\alpha)$, the number of frames used $(N)$, individual exposure duration in each filter (Exp), the filters applied (Filter), and the used telescope}.             
   \label{table_logobservations}      
   \centering          
   \begin{threeparttable}
   \begin{tabular}{ccccccccc}    
	\toprule
		Date & Time & $r$ & $\Delta$ & $\alpha$ & $N$ & Exp. & Filter & Telescope\\
		     & UT   & au  & au       &  deg     &     &  s   &        &          \\
		\midrule
		2023-07-22 & 20:14-22:10 & 3.861 & 3.548 & 15.0 & 19,19,19 & 120 & BVR & 1.3SP\\
		2023-07-30 & 20:37-22:13 & 3.779 & 3.487 & 15.4 & 10,10,10 & 200 &  VR & 61SP\\
		2023-08-24 & 20:37-21:20 & 3.514 & 3.324 & 16.7 & 5,5,5    & 200 & BVR & 61SP\\
		2023-08-25 & 21:02-22:17 & 3.503 & 3.317 & 16.8 & 10,10,10 & 200 & BVR & 61SP\\
        2023-09-01 & 17:25-18:37 & 3.430 & 3.278 & 17.1 & 15,10,10 &	90 & BVR &	1.0SIAO\\
		2023-09-05 & 21:46-00:27 & 3.384 & 3.253 & 17.3 & 16,16,16	  & 200	  & BVR	& 61SP\\
		2023-09-06 & 20:56-00:08 & 3.374 & 3.247 & 17.4 & 19,19,19	  & 200	  & BVR	& 61SP\\
		2023-09-07 & 19:26-22:49 & 3.363 & 3.242 & 17.4 & 20,20,20	  & 200	  & BVR	& 61SP\\
		2023-09-08 & 18:23-21:35 & 3.353 & 3.237 & 17.5 & 21,21,21	  & 200	  & BVR	& 61SP\\
		2023-09-09 & 18:29-20:48 & 3.342 & 3.231 & 17.5 & 20,20,20	  & 200	  & BVR	& 61SP\\
		2023-09-10 & 18:15-21:06 & 3.331 & 3.225 & 17.6 & 20,20,20	  & 200	  & BVR	& 61SP\\
		2023-09-12 & 18:31-21:01 & 3.309 & 3.214 & 17.7 & 21,21,21	  & 200	  & BVR	& 61SP\\
		2023-09-15 & 18:35-22:09 & 3.275 & 3.196 & 17.8 & 20,20,20	  & 200	  & BVR	& 61SP\\
		2023-09-17 & 20:03-20:48 & 3.254 & 3.184 & 17.9 & 12,12,12	  & 200	  & BVR	& 61SP\\
		2023-09-20 & 18:00-19:29 & 3.221 & 3.167 & 18.1 & 8,8,8	      & 240   & BVR & 61SP\\
		2023-09-25 & 18:24-22:25 & 3.165 & 3.137 & 18.3 & 20,20,20	  & 240	  & BVR	& 61SP\\
		2023-09-27 & 17:50-18:48 & 3.143 & 3.124 & 18.4 & 6,6,6	      & 240   & BVR & 61SP\\
		2023-10-10 & 14:23-17:04 & 2.998 & 3.041 & 19.0 & 37,36,36 & 30-100 &BVR&1.0SIAO\\
		2023-10-11 & 14:04-15:06 & 2.987 & 3.035 & 19.1 & 18,18,19	  & 100	  & BVR	&1.0SIAO\\
		2023-10-17 & 19:04-20:00 & 2.916 & 2.992 & 19.4 & 13,13,13	  & 200	  & BVR	& 61SP\\
		2023-10-23 & 19:26-20:41 & 2.846 & 2.947 & 19.7 & 10,10,10	  & 200	  & BVR	& 61SP\\
		2023-11-01 & 18:49-20:26 & 2.741 & 2.876 & 20.2 & 21,21,21	  & 200	  & BVR	& 61SP\\
		2023-11-04 & 17:05-18:41 & 2.706 & 2.851 & 20.3 & 24,24,24	  & 200	  & BVR	& 61SP\\
		2023-11-07 & 16:27-19:39 & 2.670 & 2.825 & 20.5 & 24,24,24	  & 200	  & BVR	& 61SP\\
		2023-11-19 & 13:53-18:37 & 2.529 & 2.717 & 21.3 & 22,22,22 &	20-30 &	BVR&1.0SIAO\\
		2023-11-20 & 14:37-15:52 & 2.517 & 2.708 & 21.4 & 30,30,30 & 30    & BVR&1.0SIAO\\
        2023-11-21 & 12:53-14:13 & 2.512 & 2.704 & 21.4 & 2           & 120   & $r^{\prime}$ &2.0HCT*\\ 
        2023-11-22 & 13:52-14:10 & 2.500 & 2.694 & 21.5 & 1           & 900   & Spectra&2.0HCT  \\
        2024-03-10 & 16:55-17:26 & 1.086 & 1.647 & 39.7 & 24          & 20    & R   &AZT8LSKU*\\
		\bottomrule
.   \end{tabular}
   \begin{tablenotes}
   \item \vspace*{-3ex} *Observations were used only for morphology and analysis of active structures.
   \end{tablenotes}
   \end{threeparttable}
   \end{table*}

\section{Results of observations}
\subsection{Morphology}
\begin{figure}
	\centering
	\includegraphics[width=.69\textwidth]{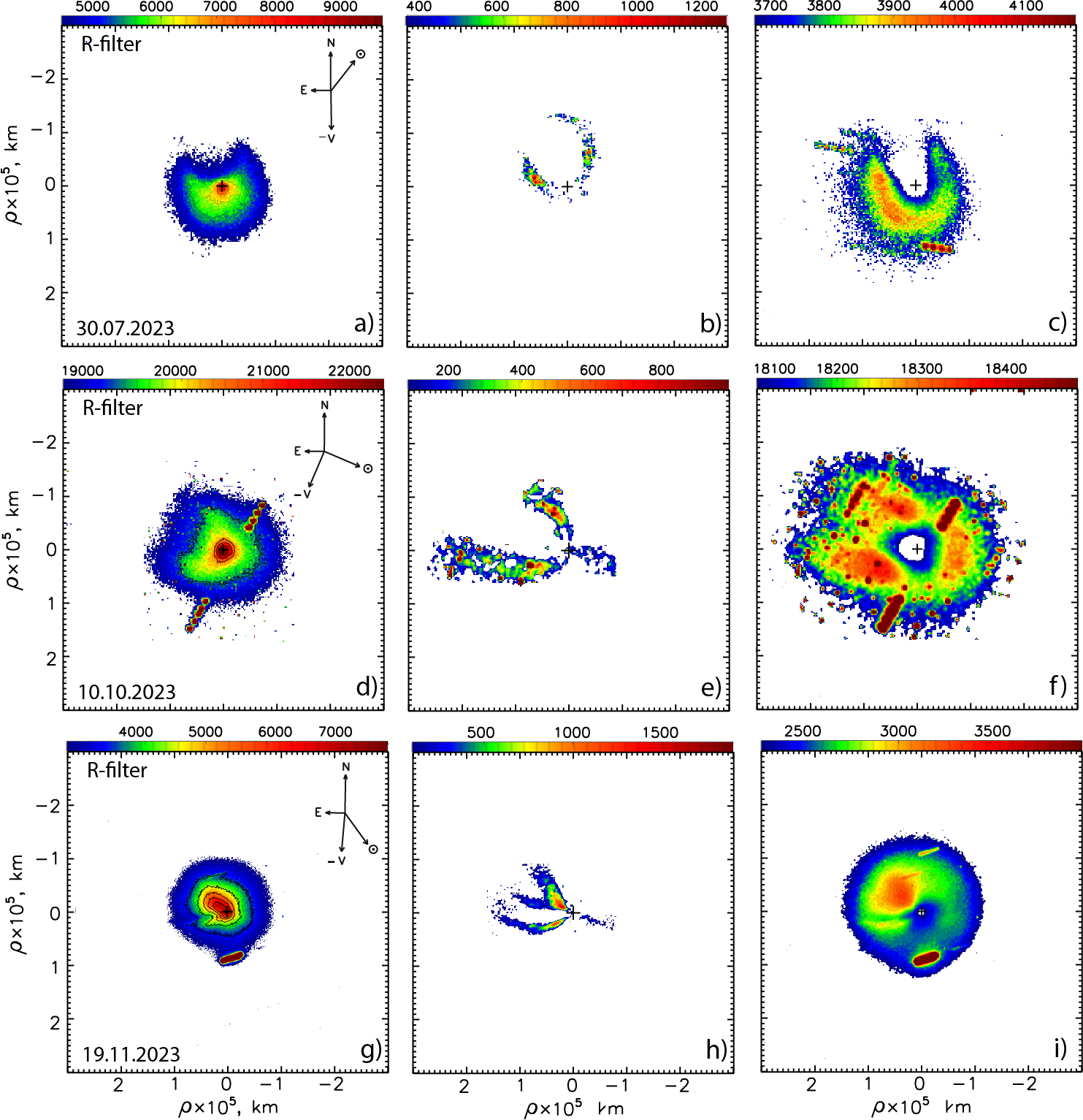}
	\caption{Intensity maps of comet 12P/Pons-Brooks in the R filter on July 30, October 10, and November 19, 2023. Panels (a), (d), and (g): Direct images of the comet. Panels (b), (e), and (h): Images processed by the rotational gradient method \citep{1984AJ.....89..571L}. Panels (c), (f), and (i): Images to which a division by the $1/\rho$ profile has been applied \citep{2014Icar..239..168S}. The colour scale does not accurately reflect the comet's absolute brightness. The arrows point toward the Sun, the north, the east, and the negative velocity vector of the comet as seen on the plane of the sky.}
	\label{FIG:Fig1_fin}
\end{figure}

\begin{figure}
	\centering
	\includegraphics[width=.49\textwidth]{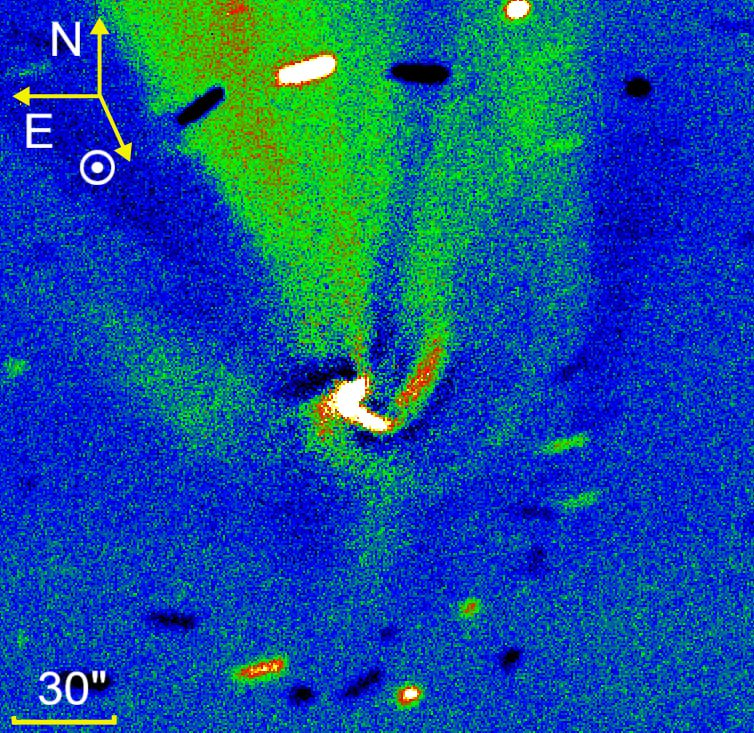}
	\caption{Co-added images of comet 12P/Pons-Brooks in the R filter to which a rotational gradient technique of \citep{1984AJ.....89..571L} is applied on March 10, 2024.}
	\label{FIG:intensity_map_March10}
\end{figure}

Comet 12P experienced several notable outbursts, providing an opportunity to study the evolution of active structures in its coma during and following these events. To investigate the coma's morphology, we constructed intensity maps for each observed date, as illustrated in Figures 1--2, which highlight the morphological changes detected throughout the observations. To identify faint features within the coma, we applied various digital filtering techniques while taking measures to minimise potential artefacts. These filters were used to individual images captured each night, as well as to coadd composite images for the same night. Structures that were not consistently visible across all images were excluded from further analysis. Additionally, we employed methods that involved shifting the image centre to account for time-dependent structural variations, ensuring that only genuine features were retained for interpretation. Such approaches, commonly used in analysing cometary comae, have proven effective in revealing dynamic structural changes (e.\,g., \citet{2010A&A...512A..60V,2015EPSC...10..622T,2017MNRAS.469S.475R,2020P&SS..18004779G}, among others). The morphology of comet 12P was analysed based on images obtained on July 30, October 10, and November 19, 2023, near the epochs of major outbursts, and March 10, 2024. According to \citet{2025AJ....169..338J}, significant outbursts occurred on July 21, October 6, November 1, 15, and 30, 2023, with additional strong events detected on December 14, 2023, and January 18, 2024. The processed images were enhanced using rotational gradient techniques of 20$^{\circ}$ \citep{1984AJ.....89..571L} and division by azimuthally averaged profiles \citep{2014Icar..239..168S}, with results for each day, July 30, October 10, and November 19, 2023, presented in the middle and right panels of Fig.\,\ref{FIG:Fig1_fin}, and for March 10, 2024 in Fig.\,\ref{FIG:intensity_map_March10}.

The morphology of the coma observed on July 30, 2023, shows an expanding, inverted umbrella-shaped structure. Panels b) and c) of Fig.\,\ref{FIG:Fig1_fin} demonstrate that the umbrella-shaped coma exhibits a well-defined spatial structure, indicating inhomogeneities in the ejected material that are likely related to the physical configuration of the active source region on the nucleus. A slightly enhanced dust density is observed on the southeastern side of the coma, and the overall elongation of the coma is strongly displaced from the exact antisolar direction.

The image acquired on October 10, 2023, following the October 6 outburst, displays a similarly horned, asymmetric coma structure, again with a higher dust concentration on the southeastern side. The umbrella-shaped coma remains elongated in the antisolar direction.

Analysis of the dust component on November 19, 2023, shows a weaker asymmetry, with the northeastern side of the coma now containing a noticeably larger dust concentration. This orientation coincides with the expected antisolar direction for the dust tail. This distribution does not necessarily reflect any preferential direction of sustained or persistent dust emission but may result from transient outburst activity.

\citet{2024MNRAS.534.1816F} also investigated the coma morphology and molecular spectral profiles of comet 12P, reporting a slight asymmetry in the gas emission profiles on November 17 -- 18, 2023. Their data indicated marginally higher molecular densities southwest of the coma than in the northeast, potentially associated with localised gas-emission activity. Their analysis of the dust component in the spectra at this time showed an increased dust density on the northeastern side of the coma, consistent with the antisolar direction of the dust tail, complicating attempts to directly correlate the dust and gas distributions.

Our observations confirm the findings of \citet{2025AJ....169..338J}, who analysed the morphological evolution of comet 12P over a wide range of heliocentric distances, from 7.89\,au (March 30, 2022) to 2.39\,au (December 1, 2023). They noted that the coma morphology changed dramatically on July 24, 2023 (3.85\,au) following an outburst, acquiring a horned appearance that disappeared by September 20, 2023, only to reappear in later observations, persisting until at least December 1, 2023 (2.39\,au). From the motion of the ejecta front, the particle velocity was estimated to be $375 \pm 38$\,m/s \citep{2025AJ....169..338J}. Subsequent images (after an outburst around July 24, 2023) revealed an expanding coma resembling an umbrella, formed by dust released from the sunward side of the nucleus and then accelerated anti-sunward under the action of radiation pressure. The ``horned'' appearance of the comet is explained as a projection effect of this dust distribution on the plane of the sky, which creates the impression of an empty region behind the nucleus.

Our images also show the ``horned'' morphology of the coma. This morphology is present in our data over an extended interval, from July 30 to November 19, 2023. Such long-term persistence of the ``horns'' cannot be explained by a single outburst. To justify this statement, we estimate the size of the larger horn in Fig.~\ref{FIG:Fig1_fin}e, which is $2.5 \times 10^{5}$~km. According to the photometric data presented by \cite{2025AJ....169..338J}, the nearest outburst occurred on October 6, 2023, four days before our observation. A simple calculation shows that, even without accounting for the angle between the line of sight and the particle velocity vector, the projected velocity in the plane of the sky would have to exceed 0.7~km/s. This value is significantly higher than expected dust-particle velocities. If one applies the velocity-estimation method of \cite{2025AJ....169..338J}, including the phase-angle correction, the resulting velocity is unrealistically high, about 2.1~km/s. The previous outburst occurred much earlier, on July 21, and therefore could not have contributed to the formation of the visible ``horns'' on October 10.
The images from October 11 and November 19, 2023, show an almost rectilinear jet structure directed approximately toward the Sun. 
One possible interpretation is that this feature originates from a near‑polar active region. This would imply that the orientation of the nucleus spin axis must be close to the position angle of this active structure. Another potential explanation is that a sunward jet could form directly from the subsolar point, where the outgassing rate reaches its maximum \citep{snodgrass201767p}. However, several arguments speak against this interpretation. First, such a jet is observed only briefly, appearing only in the images from October 11 and November 19, 2023. Second, the region of maximum outgassing is relatively broad and does not provide conditions for the formation of a narrow, well‑defined jet \cite{tubiana2019diurnal}. Third, at the observing epoch on November 19, 2023, the corresponding jet already shows a noticeable deviation from the solar direction.
A good example of how a sunward‑pointing jet can be mistakenly interpreted as originating from the subsolar region is provided by \cite{boehnhardt2016mt}. They present a table of jet position angles for comet 67P/Churyumov–Gerasimenko from August 22, 2015 to May 8, 2016. According to their data, the phase angle of jet B differed from the solar direction by no more than $8^\circ$ during the early period up to October 10, 2015. However, toward the end of the observing interval, starting around December 29, 2015, this deviation increased dramatically, reaching a maximum of $164^\circ$. The jet position angle changed smoothly over time, ruling out the possibility of misidentifying the jet at different epochs. Clearly, only a sufficiently long observational baseline allows one to unambiguously associate a sunward‑oriented jet with the subsolar region on the nucleus. These considerations led us to adopt the interpretation that the active region responsible for the observed structure is located near the pole.
Such a configuration is expected when the jet axis lies close to the rotation axis of the nucleus, and when its projection onto the plane of the sky is also close to the projected rotation axis. Enhanced gas and dust ejection from the subsolar point cannot explain this jet, as the active structure observed on November 19 is significantly offset from the solar direction. These observations allowed us to refine the orientation parameters of the rotation axis for all available observing dates.

\subsubsection{Geometric model of observed active structures}
To interpret the observed active structures, we applied the geometric model \citep{2025Icar..42516300K}. This model provides a straightforward approach to interpreting feature morphology by focusing on their shape. The model constrains the distance of jet particles from the nucleus to neglect the size-dependent acceleration due to solar radiation pressure. The geometric model primarily analyses the rotation parameters of the nucleus and the locations of active regions on the surface that serve as sources of jet structure. It is not intended to reproduce the entire coma; instead, it generates only the jet patterns produced by a limited number of active regions on the nucleus.
Using data from a single date often does not allow one to select a unique set of model parameters that reproduces the observed active structure morphology. However, applying the geometric model across multiple observation epochs enables the rejection of inconsistent parameter sets and the selection of a single set that provides an adequate description across all observing dates. The geometric model has already proven effective in interpreting complex jet structures of numerous comets, including unusual morphologies, displacement of the optocenter relative to the nucleus, and the emergence of multiple jets from a single active region \citep{rosenbush2021photometry,shubina2023photometry,voitko2025photometric,rosenbush2026comprehensive}.

\begin{figure}
	\centering
	\includegraphics[width=.5\textwidth]{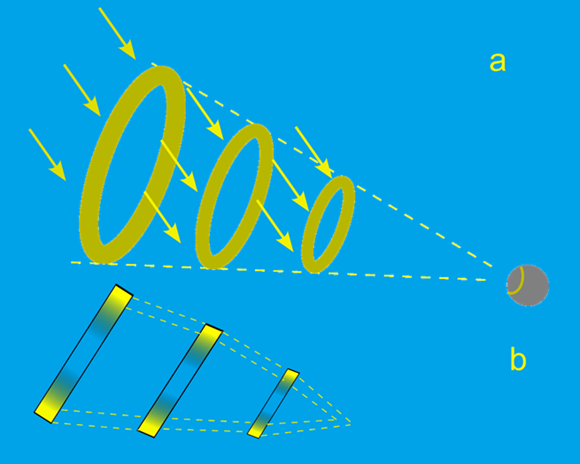}
	\caption{Schematic diagram of the formation of an active structure in the form of two ``horns'' for comet 12P/Pons-Brooks: panel (a) shows the existence of a jet structure with a cone-like geometry, and panel (b) schematically illustrates the corresponding brightness distribution of such a feature as seen by an observer.}
	\label{FIG:schematic_diagram}
\end{figure}

Interpreting the observations for this comet proved rather challenging. The main difficulty is that a single set of nuclear rotation parameters cannot account for the observed morphology across all observation dates. In addition, the unusual shape of the jet structures requires further clarification of how the model images should be interpreted. For illustration, Figure~\ref{FIG:schematic_diagram} provides a simplified schematic of how such ``horned'' structures may arise. Panel~\ref{FIG:schematic_diagram}a schematically depicts an active structure with a cone-like geometry. Such a structure may form as a result of material ejection from a confined active region on the nucleus and/or the rotation of the nucleus. Cross-sections at different distances from the nucleus are indicated, with arrows showing the viewing directions. Panel 3b illustrates the corresponding brightness distribution of such a feature as seen by an observer. The active structure brightness increases significantly toward the edge of the cone, since in this region the line of sight crosses the structure along a longer path, and therefore the apparent surface density of particles is maximised. Panel 3b also schematically indicates with dashed lines the location of the visible trace of such a feature on the coma image. Indeed, in real situations, the brightness distribution can be modulated by nucleus rotation, variations in outgassing activity, particle acceleration by solar radiation pressure, and changes in the background coma brightness. Jet structures are generally enhanced using digital filters. The most commonly applied is the rotational gradient filter \citep{1984AJ.....89..571L}, which is particularly effective at highlighting features with predominantly radial orientations, but is less sensitive to azimuthal structures. Consequently, in images processed with this filter, only the radial component of the structure is visible, appearing as two ``horns''. It is this selected structure that we used as the basis for our modelling.

For the modelling, we used the jets highlighted in panels b, e, and h from Fig.~\ref{FIG:Fig1_fin}. These images were obtained in the R filter. Spectroscopic analysis of the comet (Section 3.3)  shows that the contribution of gas emission lines within this filter does not exceed 5\,\%. Therefore, with high confidence, we can conclude that the observed jets consist predominantly of dust particles.

\begin{figure}
	\centering
	\includegraphics[width=.7\textwidth]{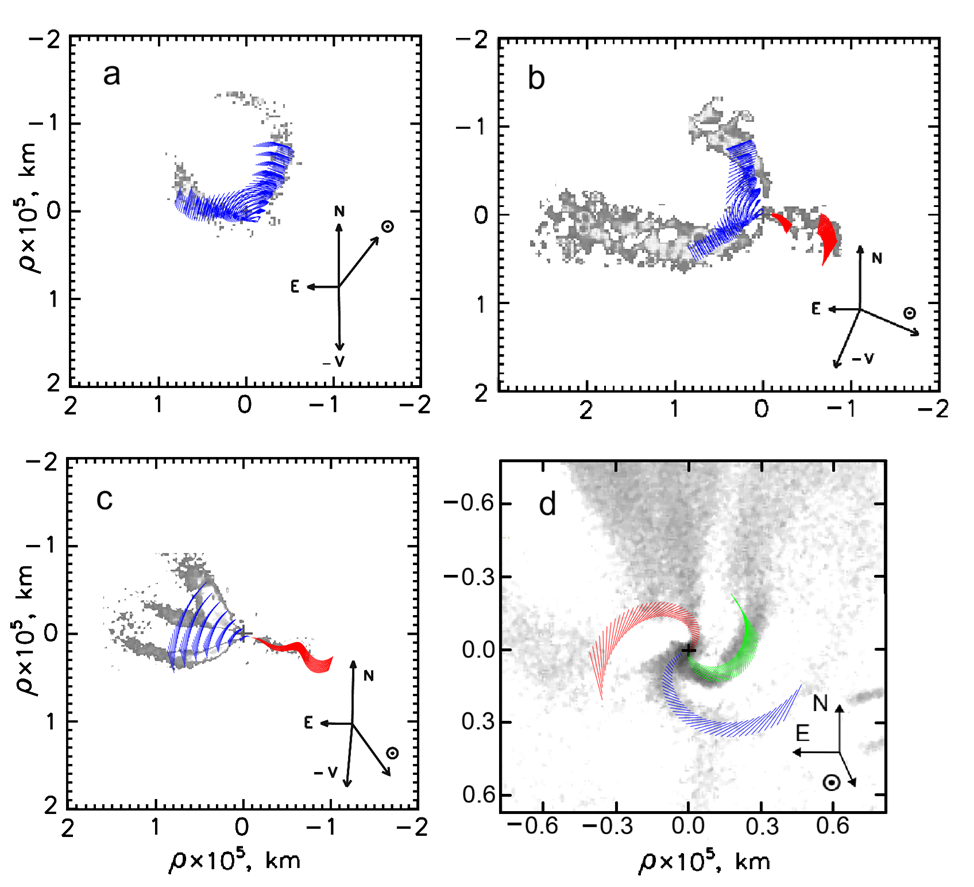}
	\caption{Application of a geometric model to interpret the structure of jet features (a -- for July 30, 2023, b -- for October 10, 2023, c -- for November 19, 2023, d -- for March 10, 2024). Active structures that come from different regions are shown in different colours.}
	\label{FIG:app_geometric_model}
\end{figure}

The geometric model parameters for the images obtained on July 30, 2023, were selected to maximise consistency with those derived for other dates. The most general feature of the images displaying the ``horned'' morphology is the presence of a relatively wide active belt in longitude at latitude $\varphi = +43^\circ \pm 5^\circ$, which produces a relatively narrow latitudinal outflow with an ejection angle of $\varepsilon = 6^\circ \pm 2^\circ$. Within this model, such a belt was represented by four discrete active regions at the same latitude, spanning $270^\circ$ in longitude. The coordinates of the positive pole of the nucleus were determined to be right ascension $\alpha = 75^\circ \pm 5^\circ$ and declination $\delta = 5^\circ \pm 5^\circ$. The latitude of the subsolar point for this date is $\varphi = +36^{\circ} \pm 6^{\circ}$. The value of the latitude of the subsolar point is given with an error due to the uncertainty in determining the orientation of the rotation axis of the nucleus. For this particular image, the rotation period of the nucleus and the particle velocity in the active structure cannot be constrained, owing to the superposition of different portions of the jet generated by the broad active zone that produces the observed structure. The model jet structure, overlaid on the observed image for July 30, 2023, is shown in Fig.\,\ref{FIG:app_geometric_model}a.

The principal difference in the observed active structure morphology on October 10, 2023, compared to the earlier date, is the presence of an almost rectilinear feature oriented toward the sun. Such a structure originates from a near-polar region, implying that the position angle of the nucleus's rotation axis must be close to the position angle of this active feature. The modelled latitude of the active area on the nucleus producing this structure is $\varphi = -80^\circ \pm 4^\circ$, with a latitudinal open angle of $\varepsilon = 10^\circ \pm 3^\circ$. In addition to the sunward-directed active structure, the ``horned'' morphology is still present, formed by a longitudinally extended active belt at latitude $\varphi = +43^\circ \pm 5^\circ$, producing a narrow latitudinal outflow with an opening angle of $\varepsilon = 6^\circ \pm 2^\circ$. The coordinates of the positive pole of the nucleus were derived as right ascension $\alpha = 20^\circ \pm 5^\circ$ and declination $\delta = 0^\circ \pm 5^\circ$. The latitude of the subsolar point for this date is $\varphi = -1^{\circ} \pm 6^{\circ}$. As in the previous case, the rotation period of the nucleus and the particle velocity within the structure cannot be constrained from this image due to the superposition of different feature components. The model jet structures overlaid on the observed structure for October 11, 2023, are shown in Fig.\,\ref{FIG:app_geometric_model}b.

On November 19, 2023, an almost rectilinear jet structure directed toward the Sun was again observed. This feature is even narrower than on the previous date, reducing the possible deviation between its position angle and the nucleus's rotation axis. The latitude of the active region on the nucleus producing this feature is $\varphi = -85^\circ \pm 3^\circ$, which, within uncertainties, coincides with the active area identified during the previous observing epoch. It is most likely the same region, though with reduced activity. This interpretation is supported by the lower contrast of the structure and its smaller latitudinal half-width of $\varepsilon = 7^\circ \pm 3^\circ$. The ``horned'' morphology is also present on this date, produced by a longitudinally extended active belt at latitude $\varphi = +43^\circ \pm 5^\circ$, with a narrow latitudinal outflow characterised by an opening angle of $\varepsilon = 8^\circ \pm 2^\circ$. The image also shows the formation of the dust tail, located between the northern and southern ``horns''. The forming tail distorts the shape of the northern horn, while the southern horn has also changed. Instead of a smooth arc at about 10$^{5}$\,km, a kink is observed, most likely associated with particle acceleration by solar radiation pressure or a rapid increase in particle velocity. All these effects are consistent with the decreasing heliocentric distance. The coordinates of the positive pole of the nucleus were determined as right ascension $\alpha = 355^\circ \pm 5^\circ$ and declination $\delta = 25^\circ \pm 5^\circ$. The latitude of the subsolar point for this date is $\varphi = -37^{\circ} \pm 6^{\circ}$. As in the previous cases, the rotation period of the nucleus and the particle velocity in the active structure cannot be constrained from this image due to the superposition of its different components. The model jet structures overlaid on the observed structure for November 19, 2023, are shown in Fig.\,\ref{FIG:app_geometric_model}c.

For March 10, 2024, the active structure morphology also changed dramatically. This transformation is associated with the comet's altered orbital position and the rotation axis's changed orientation. The illumination conditions of the nucleus surface evolved in such a way that the broad active belt at latitude $\varphi = +43^\circ \pm 5^\circ$ remained permanently in shadow and therefore exhibited no activity. The region near the negative pole also appeared inactive, most likely due to reduced activity already noted on November 19, 2023. At this epoch, three distinct active structures with a spiral-like morphology were observed. Such a configuration imposes constraints on the rotation sense of the nucleus and indicates that the comet's negative pole was oriented toward the observer. The first feature, shown in blue in Figure\,\ref{FIG:app_geometric_model}d, originates near the equator at latitude $\varphi = 0^\circ \pm 5^\circ$ and longitude $\lambda = 0^\circ$. Its opening angle is $\varepsilon = 10^\circ \pm 4^\circ$. The longitudes of the remaining active regions are given relative to this first region. The second feature, marked in red in Figure \,\ref{FIG:app_geometric_model}d, has an opening angle of $\varepsilon = 12^\circ \pm  4^\circ$ and arises from a region at latitude $\varphi = -30^\circ \pm 5^\circ$ and longitude $\lambda = 145^\circ \pm 10^\circ$. The third feature, shown in green in Figure\,\ref{FIG:app_geometric_model}d, originates at latitude $\varphi = -55^\circ \pm 5^\circ$ and longitude $\lambda = 290^\circ \pm 10^\circ$, with an opening angle of $\varepsilon = 12^\circ \pm 4^\circ$. The coordinates of the positive pole of the nucleus were derived as right ascension $\alpha = 15^\circ \pm 5^\circ$ and declination $\delta = 25^\circ \pm 5^\circ$. The latitude of the subsolar point for this date is $\varphi = -63^{\circ} \pm 6^{\circ}$.

In addition to depending on the orientation of the rotation axis and the relative geometry of the observer and the nucleus, the morphology of the active structure observed in the coma is governed by a characteristic scale equal to the product of the comet’s rotation period and the mean particle velocity in the jet. Based on our data, we cannot determine the rotation period of the nucleus; therefore, we provide only the value of this characteristic scale, which is $1.24 \times 10^{5}$~km. Assuming a nucleus rotation period of 57\,hours \citep{2024ATel16508....1K}, this yields a particle velocity of about 0.6\,km\,s$^{-1}$. If we adopt a typical range of dust-particle velocities of 200--300~m/s, the corresponding rotation period would be 114--171 hours. The value of 57 hours was obtained from earlier observations conducted between January 31 and February 16, 2024. Our determinations refer to March 10, 2024. If the nucleus of comet 12P exhibits unstable rotation, a substantial change in the rotation period over a short timescale is possible. Similar behaviour has been documented for comet 41P/Tuttle–Giacobini–Kres\'{a}k \citep{2018Natur.553..186B,2019AJ....157..108S}.

\subsection{Dust activity and colours}
\begin{figure}
	\centering
	\includegraphics[width=.7\textwidth]{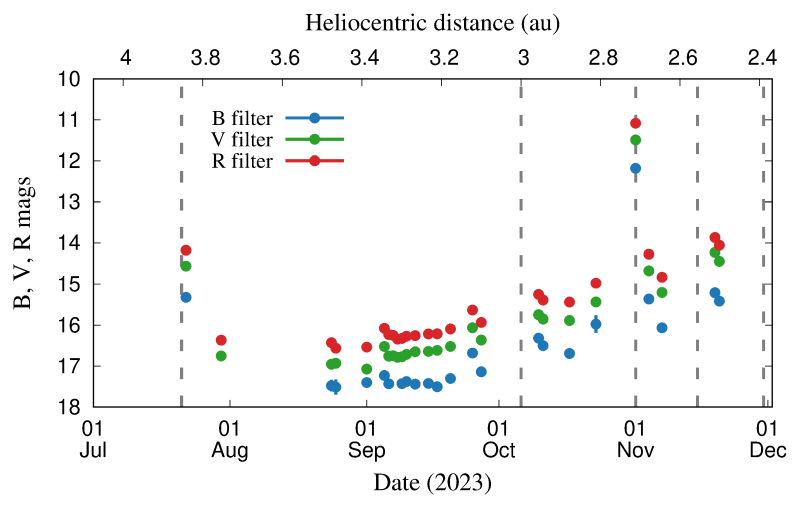}
	\caption{The distribution of the apparent magnitudes in B, V, and R bands of comet 12P/Pons-Brooks throughout the entire observation period. The figure shows changes in magnitude over time and with heliocentric distance. Vertical dashed lines indicate dates of outbursts taken from the literature (see Section 3.1 for more details).}
	\label{FIG:BVRmags_vs_date}
\end{figure}

\begin{figure*}
	\centering
	\includegraphics[width=.7\textwidth]{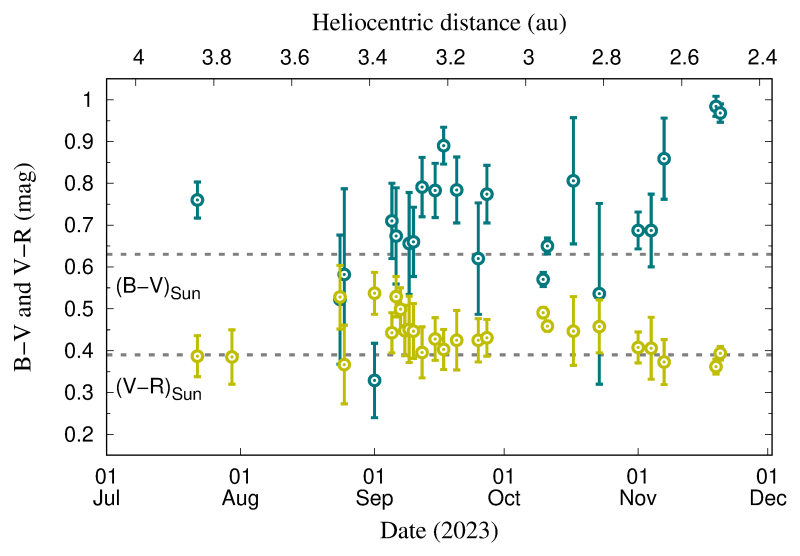}
	\caption{The change of the B-V (darker green points) and V-R (light green points) colour of comet 12P/Pons-Brooks throughout the entire observation period and heliocentric distances. Horizontal lines indicate solar colours in appropriate filters.}
	\label{FIG:B-V_vs_date}
\end{figure*}

To determine the comet's absolute magnitude, $Af\rho$ parameter \citep{1984AJ.....89..579A}, photometric colour, and normalised reflectivity gradient, we calculated its apparent magnitudes. In the context of the comet's gradual brightness changes due to its approach to perihelion (see Fig.\,\ref{FIG:BVRmags_vs_date}, \ref{FIG:B-V_vs_date}, respectively for apparent magnitudes and colours), we also observed two outbursts where its magnitude sharply increased by several units, specifically on July 22 and November~1, 2023.

The absolute magnitude was derived by adjusting the apparent magnitude to account for the unit heliocentric and geocentric distances.
The formula is as follows, where $m$ represents the comet's apparent magnitude has no $\Phi(\alpha)$ as the phase function. However, accurately estimating the phase function is challenging due to factors such as the comet's nucleus rotation, seasonal variations, and outburst activity \citep{1998Icar..132..397S}. While different researchers use different phase functions, we opted to focus first on the absolute magnitude to ensure a more universally applicable calculation.
\begin{equation}
H(1, 1) = m - 5\log(r\Delta).
\end{equation}

Next, we calculated the $Af\rho$ parameter, which serves as a measure of a comet's dust activity level. This parameter is obtained by multiplying the albedo $(A)$, the filling factor $(f)$, which represents the ratio of the area covered by dust particles within the aperture to the total aperture area, and the linear aperture radius $(\rho)$ at the comet \citep{1984AJ.....89..579A}. One of its key advantages is that it is unaffected by observation conditions or the instruments used \citep{2012Icar..221..721F}. Additionally, this parameter is relatively straightforward to compute, as demonstrated by \citet{2009A&A...502..355M}:
\begin{equation}
Af\rho = 4 r^2 \Delta^2 10^{0.4(m_{\mathrm{Sun}} - m)}/\rho ,
\end{equation}
where $m_{\mathrm{Sun}}$ is the solar magnitude in the respective filter, which we took from \citet{2018ApJS..236...47W}. Additionally, we have adjusted the $Af\rho$ parameter to $0^\circ$ phase angle using the phase function, $S(\alpha)$, defined by \citet{2011AJ....141..177S} as
\begin{equation}
A(0)f\rho = A(\alpha)f\rho / S(\alpha).
\end{equation}

\begin{figure}
	\centering
	\includegraphics[width=.45\textwidth]{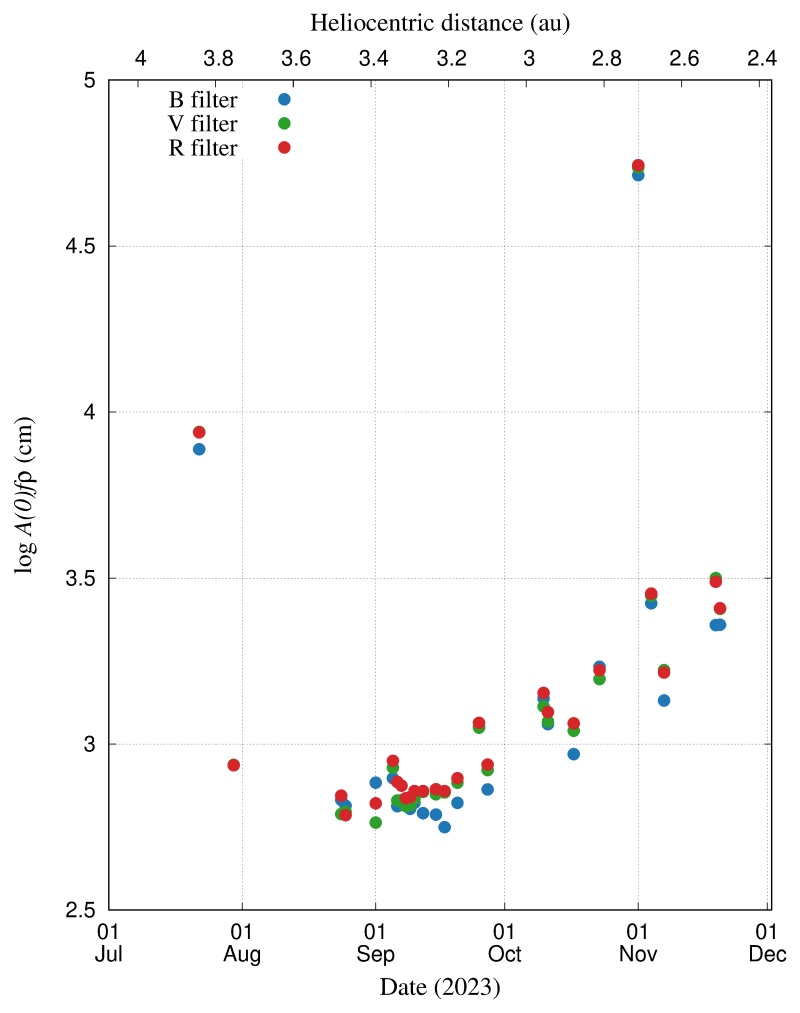}
	\caption{The change of the $\log A(0)f\rho$ parameter from B, V, and R filters of comet 12P/Pons-Brooks throughout the entire observation period and heliocentric distance.}
	\label{FIG:Afrho_vs_date}
\end{figure}

Apparent magnitudes, absolute magnitudes, and the $Af\rho$ parameter obtained in the R filter and adjusted at $0^\circ$ phase angle are listed in Table\,\ref{table_mags_afrho}\footnote{We have provided here the measurements obtained in the R band. Calculations made in B and V are available via reasonable request to the authors.}. Also, Figure~\ref{FIG:Afrho_vs_date} demonstrates $A(0)f\rho$ parameter in all filters used. 

\begin{table}
	\caption{Apparent, absolute magnitudes, aperture radii, and the reduced $A(0)f\rho$ parameter in the R filter for the comet 12P.}
	\label{table_mags_afrho}
	\begin{tabular}{ccccr}
    \toprule
Data & $m_\mathrm{R}$, mag & $H_\mathrm{R}$, mag & $\rho$, km & $A(0)f\rho$, cm\\
\midrule
2023-07-22  &  $14.18 \pm 0.04$ & 8.49  &  9700 & $8686 \pm 344$\\
2023-07-30  &  $16.37 \pm 0.05$ & 10.77 & 12160 & $863 \pm 36$\\
2023-08-24  &  $16.43 \pm 0.04$ & 11.09 & 11600 & $699 \pm 26$\\
2023-08-25  &  $16.56 \pm 0.04$ & 11.24 & 11570 & $611 \pm 21$\\
2023-09-01  &  $16.53 \pm 0.03$ & 11.28 & 10330 & $663 \pm 16$\\ 
2023-09-05  &  $16.07 \pm 0.02$ & 10.87 & 11350 & $890 \pm 20$\\
2023-09-06  &  $16.23 \pm 0.03$ & 11.03 & 11330 & $769 \pm 20$\\
2023-09-07  &  $16.25 \pm 0.04$ & 11.06 & 11310 & $749 \pm 23$\\
2023-09-08  &  $16.34 \pm 0.02$ & 11.16 & 11290 & $687 \pm 14$\\
2023-09-09  &  $16.32 \pm 0.06$ & 11.15 & 11270 & $692 \pm 40$\\
2023-09-10  &  $16.27 \pm 0.05$ & 11.11 & 11250 & $722 \pm 31$\\
2023-09-12  &  $16.25 \pm 0.04$ & 11.12 & 11210 & $722 \pm 29$\\
2023-09-15  &  $16.21 \pm 0.04$ & 11.11 & 11150 & $731 \pm 29$\\
2023-09-17  &  $16.21 \pm 0.03$ & 11.13 & 11110 & $722 \pm 18$\\
2023-09-20  &  $16.09 \pm 0.03$ & 11.05 & 11050 & $789 \pm 22$\\
2023-09-25  &  $15.64 \pm 0.02$ & 10.65 & 10940 & $1158 \pm 24$\\
2023-09-27  &  $15.93 \pm 0.03$ & 10.97 & 10900 & $867 \pm 24$\\
2023-10-10  &  $15.25 \pm 0.01$ & 10.46 & 10850 & $1425 \pm 10$\\
2023-10-11  &  $15.39 \pm 0.01$ & 10.60 & 10830 & $1248 \pm 10$\\
2023-10-17  &  $15.44 \pm 0.02$ & 10.74 & 10440 & $1155 \pm 23$\\
2023-10-23  &  $14.98 \pm 0.04$ & 10.36 & 10280 & $1670 \pm 54$\\
2023-11-01  &  $11.08 \pm 0.03$ & 6.60  & 10040 & $55389 \pm 1373$\\
2023-11-04  &  $14.27 \pm 0.06$ & 9.84	&  9950 & $2840 \pm 143$\\
2023-11-07  &  $14.83 \pm 0.04$ & 10.45 &  9860 & $1642 \pm 53$\\
2023-11-19  &  $13.87 \pm 0.01$ & 9.68  & 10840 & $3081 \pm 38$\\
2023-11-20  &  $14.05 \pm 0.01$ & 8.88  & 10800 & $2565 \pm 24$\\
\bottomrule
	\end{tabular}
\end{table}

\begin{figure}
	\centering
	\includegraphics[width=.49\textwidth]{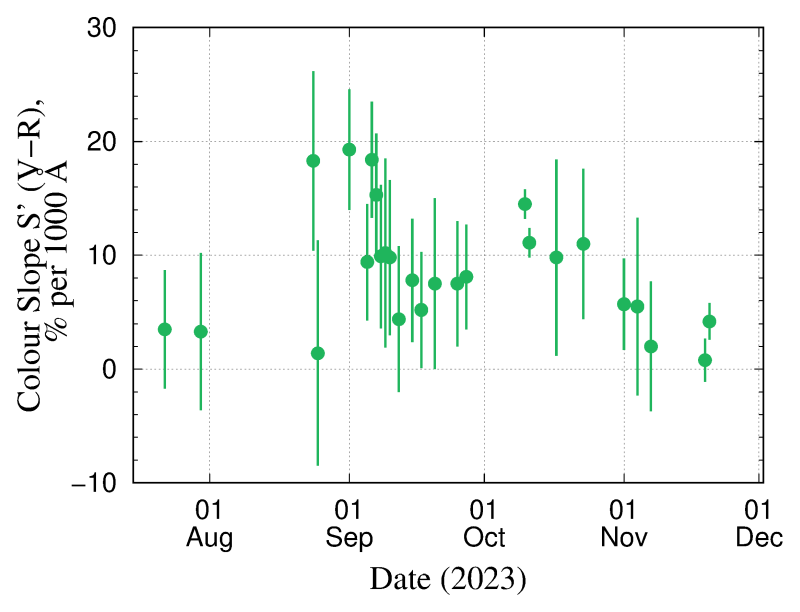}
	\caption{The change of the V-R colour slope $S^\prime$ of comet 12P/Pons-Brooks throughout the entire observation period.}
	\label{FIG:Slope_vs_date}
\end{figure}

Some authors prefer to characterise the cometary colour in terms of the so-called colour slope \citep{1984AJ.....89..579A}. This characteristic is also referred to as the reflectivity gradient, which can be measured in per cent reddening per 1000~\AA~ and is defined as follows \citep{1997EM&P...79..221J}:
\begin{equation}
S^\prime (\lambda_1, \lambda_2) = \left( \frac{2000}{\lambda_2 - \lambda_1} \right) \frac{10^{0.4{\Delta m}} - 1}{10^{0.4{\Delta m}} + 1} ,
\end{equation}
where $\Delta m$ marks the true colour index of the comet, i.\,e. the observed colour reduced for the solar colour: {\bf $(V-R)_\mathrm{comet} - (V-R)_\mathrm{Sun}$}. The resulted colour slop, $S^{\prime}(V-R)$ is presented in Figure~\ref{FIG:Slope_vs_date}. 

\subsection{Spectra}
\begin{figure}
	\centering
	\includegraphics[width=.49\textwidth]{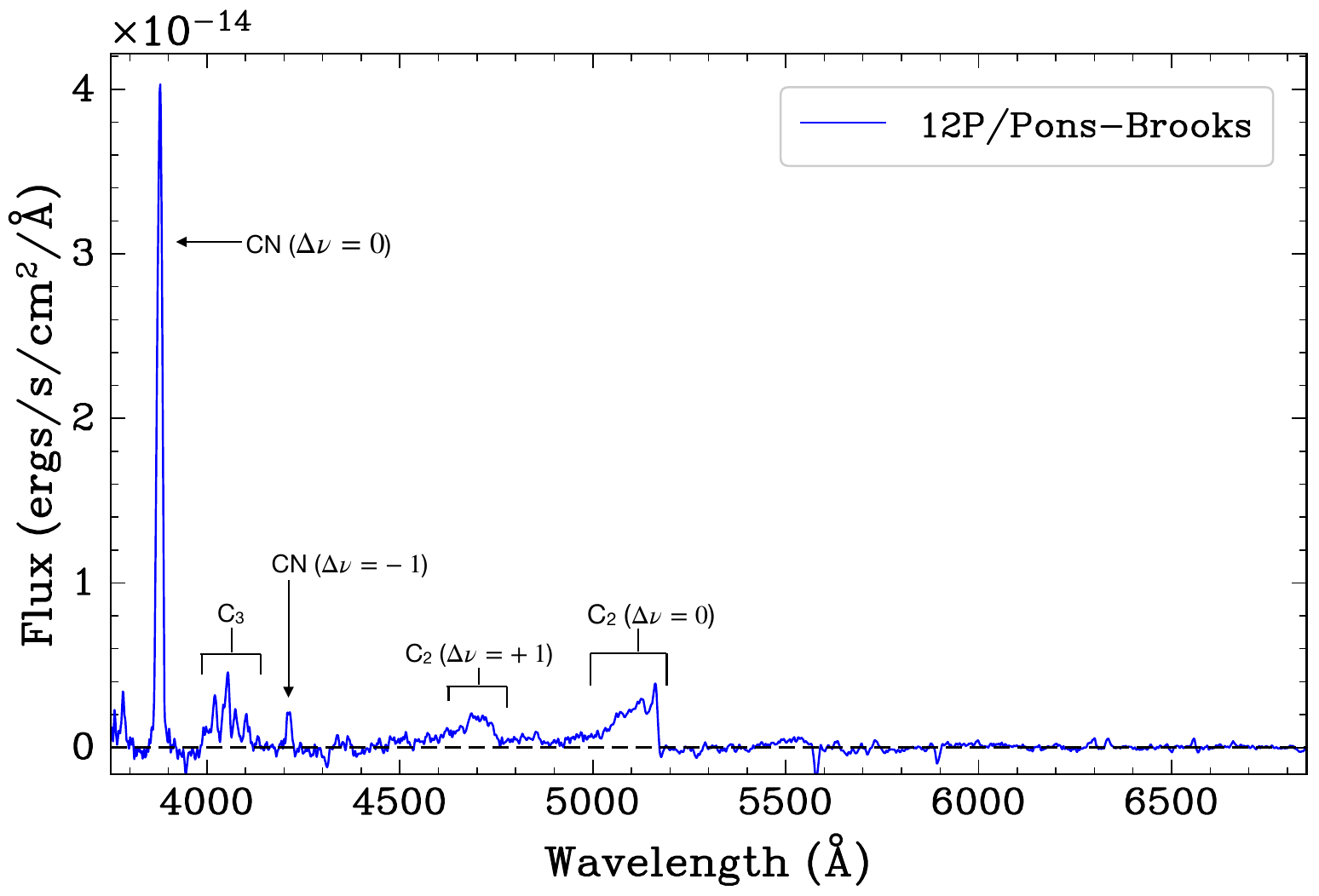}
	\caption{Emission spectrum of comet 12P with detected molecular features. 900-sec exposure spectrum was obtained on November 22, 2023, with the 2.0HCT (details in Sec. 2.5).}
	\label{FIG:Spectra}
\end{figure}

The flux-calibrated continuum-subtracted emission spectrum extracted within a 30$^{\prime\prime} \times 1.92^{\prime\prime}$ aperture centred on the comet is shown in Fig.\,\ref{FIG:Spectra}.

Emissions corresponding to CN, C$_{2}$ and C$_{3}$ were detected in the optical spectra of comet 12P. The flux of these molecules within their corresponding wavelength range was obtained from the extracted optical spectrum. Following the method described in Section 2.3 of \cite{2025A&A...701A.161A}, the Haser factor for each molecule was computed to convert the observed flux to total flux. The Haser model is a simple model that assumes a spherically symmetric coma spreading around a spherical nucleus, in which the parent species are disintegrated into daughter species only through a first-order photodissociation reaction. Furthermore, the total flux of each molecule was converted to their respective column densities \citep[see][for details]{2021MNRAS.502.3491A} and the production rates were computed with the help of the fluorescence efficiencies and daughter scale length adopted from \cite{1995Icar..118..223A}. Although the fluorescence efficiencies of C$_{2}$ and C$_{3}$ were taken from \cite{1995Icar..118..223A} and scaled to $r^{-2}_\mathrm{h}$, the efficiency of CN was obtained by performing a double interpolation in the table provided by \cite{2010AJ....140..973S}, which contains the g factor of CN for different heliocentric distances and velocities. The gas production rates, CN$ = (1.56 \pm 0.02) \times 10^{26}$\,mol/sec; C$_{2} = (1.59 \pm 0.03) \times 10^{26}$\,mol/sec and C$_3 = (2.07 \pm 0.16) \times 10^{25}$\,mol/sec, are in very good agreement with the other studies of the comet around the same epoch \citep{2024MNRAS.534.1816F,2025epsc.conf.1646V}. The production rate ratios, $\mathrm{C_2/CN} = 1.02 \pm 0.02$ and $\mathrm{C_3/CN} = 0.133 \pm 0.01$, classify the comet to be of typical carbon chain composition as per the classification provided by \cite{1995Icar..118..223A}.

\section{Discussion and Conclusion}
Our photometric observations show that between July 2023 and March 2024, the dust activity of comet 12P underwent significant evolution, accompanied by changes in brightness, the $Af\rho$ parameter, and colour indices. As the comet approached the Sun, several brightness enhancements were recorded, each corresponding to a sharp increase in dust activity. A particularly substantial rise in $Af\rho$ was observed both in late September and during the major outburst on November 1, 2023. Such values greatly exceed those characteristic of the comet’s quiescent evolution and indicate the release of large amounts of fresh dust. Comparing these variations with gas production data from other studies shows that the behaviour of dust and gas in comet 12P differed markedly. According to TRAPPIST measurements, during the same periods when dust activity increased by several factors, the production rates of CN and C$_2$ rose only moderately. This indicates that the activity outbursts were driven primarily by enhanced dust outbursts without a proportional increase in gas emission. Similar discrepancies between gas and dust have been observed in other comets with long-lasting activity at large heliocentric distances \citep[e.\,g.,][]{1997P&SS...45..455S,2003A&A...399..763L} and may reflect structural properties of near-surface layers or fragmentation processes exposing regions with low volatile abundance.
Changes in colours also support variability in the composition and size distribution of dust particles. The measured B--V colour index exhibits noticeable variations relative to the solar colour (Fig.~\ref{FIG:B-V_vs_date}). On August 24--25, September 1, and October 10 and 23, the colour became bluer than the Sun. During the remaining observing nights, the colour was either close to the solar value or indicated reddening; however, after November 1, a moderate increase in reddening was observed. The V--R colour index (Fig.~\ref{FIG:B-V_vs_date}) remained close to the solar value throughout most of the observing period, except on August 24, September 1 and 5--9, and October 10, 11, 17, and 23, when moderate reddening was detected. Nevertheless, the derived colour slope $S^{\prime}(V-R)$ (Fig.~\ref{FIG:Slope_vs_date}) indicates that the overall behaviour of the dust colour is redder than the Sun. Only on August 25 and November 20 did the dust exhibit an approximately neutral (solar-like) colour. All other measurements show dust reddening of 5--20\,\% per 1000~\AA. Notably, studies of comet 21P/Giacobini-Zinner also revealed temporal colour changes of dust grains—from redder to bluer than the Sun \citep[e.\,g.,][]{1997P&SS...45..455S,2003A&A...399..763L}. One possible explanation for such colour variations is a change in the gas production rate; however, this hypothesis has not yet received sufficient observational support \citep{1997P&SS...45..455S,2003A&A...399..763L}. In the case of comet 12P, \cite{2023ATel16282....1J} reported that during the outbursts on September 22--25 and October 3--7, the dust production rate increased by factors of about 2 and 6, respectively, relative to nearby dates, while the gas production rate showed only minimal variations during the same periods. Our measured values are consistent with previous results for comet 12P and agree well with reddening estimates reported for other short-period comets. As shown by \cite{2005MNRAS.358..641L}, the majority of Jupiter family comets exhibit reflectivity gradients of 0--13\,\% per 1000~\AA. In particular, for comet 21P/Giacobini-Zinner, a dust reddening of approximately 15\% per 1000~\AA~ was reported by \cite{1987A&A...187..531S}, while \cite{2003A&A...399..763L} found values in the range 13--29\,\% per 1000~\AA. For short-period comets 32P/Comas Sola, 43P/Wolf-Harrington, 46P/Wirtanen, and 56P/Slaughter-Burnham, dust reddening values between 0.34 and 10\,\% per 1000~\AA~ have been reported in \cite{2011A&A...525A..62M, 2005MNRAS.358..641L, 1998A&A...335L..25L}. A persistent reddening, indicated by the normalised spectral reflectivity gradient, is typical of cometary dust composed of a mixture of silicate and carbonaceous particles, which scatter light more efficiently at longer wavelengths. Such dust colours may reflect either a higher fraction of large particles in the coma of comet 12P (an abundance of large particles in the coma may also produce a redder colour) or the release of fresh, less solar-processed material during cometary outbursts. Taken together, these results indicate that during the observation period, comet 12P underwent recurrent episodes of enhanced dust activity, likely triggered by the exposure of new active regions or mechanical disruption of near-surface layers. The absence of a direct proportional increase in gas activity suggests that these outbursts were predominantly dust-driven rather than driven by vigorous volatile sublimation.

Emission features of CN, C$_2$, and C$_3$ were identified in the spectrum of comet 12P/Pons–Brooks, and their production rates were derived using Haser's model \citep{Haser1957} with standard fluorescence efficiencies and photodissociation scale lengths. The resulting values of gas production rates are in good agreement with observations other short-period comets obtained at similar heliocentric distances \citep{2011Icar..213..280L}. The C$_2$/CN and C$_3$/CN ratios indicate a typical carbon–chain composition characteristic of most short-periodic comets. Particularly, C$_2$/CN ratio is slightly lower than the averaged typical value provided by \cite{1995Icar..118..223A}, but it is prominently higher than the depletion limit. C$_3$/CN ratio is higher than the typical one presented in \cite{1995Icar..118..223A}. This suggests that the enhanced activity of 12P is not accompanied by anomalous volatile chemistry and is more likely driven by physical processes affecting the near-surface layers. Moreover, our calculations of rates' ratios within errors are close to the respective values obtained for comet 1P/Halley at the same heliocentric distance \citep{1995Icar..118..223A}. \cite{2026A&A...705A..89V} provided gas production measurements for 12P in 2023-2024 apparition. Our results for C$_{2}$ and CN gas production are consistent with their data within the errors, while the estimation for c$_{3}$ is insignificantly lower. Nevertheless, both carbon-chain composition computations, ours and those provided in \cite{2026A&A...705A..89V}, yield almost identical values within errors.

During the 2023–2024 observing campaign, comet 12P exhibited a unique ``horned'' morphology. The persistence of this structure over an extended period makes it unlikely that outbursts alone are responsible for ejecting material in a way that could produce such a morphology. A more plausible explanation is that this phenomenon is linked to intrinsic properties of the cometary nucleus. We propose an alternative interpretation for the origin of this structure. A possible explanation is the presence of a highly elongated active region extending over a substantial portion of the nucleus. Material released from such a region, when projected onto the plane of the sky, can naturally produce a stable structure of this type. This scenario does not exclude the role of outbursts, which may trigger or enhance the long‑lasting activity of an elongated active area, for example, by removing a surface layer and exposing deeper, volatile‑rich material.

However, this interpretation requires further consideration of how such an extended active region might form. One realistic possibility is that the nucleus has a bilobate shape. In the case of comet 67P/Churyumov-Gerasimenko, such a structure was observed, and the junction between the two lobes was found to be associated with enhanced activity \citep{2017MNRAS.469S.741P,2015Sci...347a1044S,2023Icar..40115566C,2018MNRAS.479.1555F,2015Natur.526..402M}.

\begin{figure}
	\centering
	\includegraphics[width=.49\textwidth]{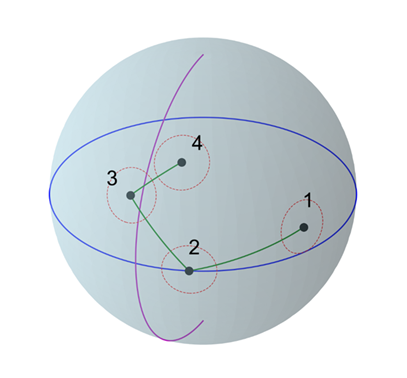}
	\caption{Change in the orientation of the rotation axis of the nucleus of comet 12P/Pons-Brooks over the entire observation period: 1 -- for July 30, 2023, 2 -- for October 11, 2023, 3 -- for November 19, 2023, 4 -- for March 10, 2024. The blue circle is the celestial equator, the magenta semicircle is the meridian of the vernal equinox, and the dotted line indicates possible errors in determining the position of the rotation axis.}
	\label{FIG:change-orient}
\end{figure}

Another potential origin for an extended active region is the formation of a large surface fracture caused by the removal of a surface layer during an outburst, by progressive surface degradation, or by the impact of a meteoroid.
Another noteworthy characteristic of this comet may be a change in the orientation of its rotation axis over a relatively short interval. For a massive body such as a cometary nucleus, changes in the spin‑axis orientation can, in principle, occur due to torques produced by outgassing. However, the expected rate of change is usually very low, so for most comets the spin‑axis orientation remains nearly constant over a single perihelion passage. In the case of comet 12P, by contrast, we observe morphological variations that can be attributed to changes of several tens of degrees in the spin‑axis orientation over the course of only about one month  (Fig.\ref{FIG:change-orient}).

According to the classical Euler problem of rigid‑body dynamics, such chaotic rotation becomes possible when the spin axis lies close to the intermediate principal moment of inertia \citep[see, e.\,g.,][]{palais2009disorienting}. In this configuration, even relatively small torques can induce substantial changes in the orientation of the rotation axis. At the same time, comet 12P exhibited a series of outbursts, which could provide the necessary torques to trigger such variations in the spin‑axis orientation. This behaviour is not unique among small bodies: examples of non‑principal‑axis rotation have already been documented for the first interstellar asteroid 1I/'Oumuamua \citep{fraser2018tumbling} and for comet 41P/Tuttle–Giacobini–Kres\'{a}k \citep{2019AJ....157..108S, 2018Natur.553..186B}.

Our analysis of comet 12P during its 2023--2024 apparition revealed several noteworthy features. The persistent ``horned'' morphology observed from late July to mid‑November 2023 could have been produced by an extended active region covering a substantial fraction of the nucleus, potentially reflecting a bilobate shape or a large surface fracture. Our modelling of the activity at different epochs indicates that different regions of the nucleus became active over time: a belt at latitude  $\varphi \approx -43^\circ$ dominated the formation of the horn‑like structure in 2023; near‑polar sources generated jets directed almost sunward in October–November; and by March 2024 the geometry between the observer, the Sun, and the nucleus favoured the appearance of three spiral‑like jet structures.

The observed morphological changes suggest that the nucleus may undergo large and rapid variations in its spin‑axis orientation on the order of several tens of degrees within roughly one month. Such behaviour is consistent with the possibility that comet 12P is in a chaotic rotational state. However, a definitive conclusion requires additional observational material spanning a longer time interval. Taken together, these results highlight that comet 12P is an exceptionally unusual object. Its persistent and atypical coma morphology, combined with its unusual outburst activity, distinguishes it from most comets studied to date and makes it an important target for continued observational and theoretical investigation.

\section*{Acknowledgements} The research of MH, OI, OSh is supported by the Slovak Academy of Sciences (grant Vega No.\,2/0067/26) and by the Slovak Research and Development Agency under Contracts No. APVV-24-0076 and No. APVV-19-0072. The research by VK funded by the Deutsche Forschungsgemeinschaft (DFG, German Research Foundation) – KL 3895/1-1. AK acknowledges support from the Wallonia-Brussels International (WBI) grant. MS acknowledges the financial support by the DST, Government of India, under the Women Scientist Scheme (PH) project reference number SR/WOS-A/PM-17/2019. MS also thanks the staff of IAO, Hanle, and CREST, Hosakote, that made these observations possible. The facilities at IAO and CREST are operated by the Indian Institute of Astrophysics, Bangalore. The research of AK was supported by the project of the Ministry of Education and Science of Ukraine No. 0124U001304.

\printcredits

\bibliographystyle{cas-model2-names}

\bibliography{references_12P}

\end{document}